\documentclass[a4paper]{article}
\usepackage{graphicx}
\usepackage{amssymb}
\usepackage{amsmath}

\newcommand{\EUV}{{\epsilon}}
\newcommand{\VOne}{\mathbb{I}}
\newcommand{\Polyn}[1]{{\cal #1}}
\newcommand{\Set}[1]{\{#1\}}

\newcommand{\SetIn}[2]{\{#1 \;|\; #2 \}}
\newcommand{\SetReal}{\mathbb{R}}

\begin{document}

\title{Sector decomposition via computational geometry}


\author{Toshiaki Kaneko %
        \\
        High Energy Accelerator Research Organization (KEK)\\
        1-1 Oho, Tsukuba, Ibaraki 305-0801, Japan \\
      \and
        Takahiro Ueda\\
        Graduate School of Pure and Applied Sciences,
        University of Tsukuba, \\
        1-1-1 Tennodai, Tsukuba, Ibaraki 305-8577, Japan\\
       }

\maketitle

\begin{abstract}{
    A non-iterative method is presented for the factorization step of
          sector decomposition method, which separates
          infrared divergent part from loop integration.
          This method is based on a classification of asymptotic
          behavior of polynomials.
          The problem is converted to ones for convex body
          in Euclidean space.
          They are solved with algorithms developed in 
          computational geometry.
          A test implementation shows that this method produces
          less number of decomposed sectors than usual
          iterative sector decompositions.
}
\end{abstract}

\vspace{2em}
{\small
Talk given at
\textsl{13th International Workshop on 
Advanced Computing and Analysis Techniques in Physics Research},
February 26, 2010,
Jaipur, India
}

\section{Introduction}

In the calculation of Feynman amplitude with massless particles, 
one has to regulate infrared divergences (IR), which 
cancel out among loop corrections and real emission processes.
In perturbative QCD, this divergences are regulated by $D$-dimensional
method.
Divergent part is expressed as poles in terms of
$\EUV = (4-D)/2$.
Sector decomposition method developed in Refs.~%
\cite{phys:Binoth-Heinrich-2000,phys:Binoth-Heinrich-2004,%
      phys:Binoth-Heinrich-2004a,phys:Heinrich-2008}
are widely used for separating IR divergences.

As a simple example of separation of infrared divergences,
let us consider the following one-dimensional integral for $\EUV < 0$
with a regular function $f$ such that $f(0)$ is a non-zero
finite value:
\begin{equation}
\begin{split}
I &= \int_0^1 dx \; x^{-1-\EUV} f(x)
\\
  &= \int_0^1 dx \; x^{-1-\EUV} f(0) 
   + \int_0^1 dx \; x^{-1-\EUV} \left( f(x) - f(0) \right)
\\
  &= - \frac{f(0)}{\EUV}
   + \int_0^1 dx\; \left(x^{-\EUV} f'(0) + \frac{1}{2} x^{1-\EUV} f''(0)
                        + ... \right) .
\end{split}
\end{equation}
When $\EUV=0$,
the factor $x^{-1-\EUV}$ produces a logarithmic divergence 
by the integration around $x \sim 0$.
While for $\EUV < 0$, the divergence is regularized and converted
to a pole of $\EUV$.
The coefficient of the pole is $f(0)$, which is the first term
of the Taylor expansion of $f(x)$ around $x=0$.
The rest of the integration becomes finite for $\EUV\rightarrow 0$.
This example shows that, 
when the singular part of the integrand is factored out,
one can separate divergent part as poles in terms of $\EUV$.

For the general loop integrations, it will also be possible to
separate IR divergence, if the singular part is factored out.
Although the factorization of a multi-variate polynomial
is not a trivial problem,
it is solved by the method of sector decomposition.
This method shows that
when the integration domain is properly decomposed into sectors and 
selecting good variables in each sector,
the divergent part is factored out.
This method provides at the same time a practical procedure to find such
sectors and variables.
Then we can obtain the coefficients of the poles
of $\EUV$ and the finite part as a Laurent series in terms of $\EUV$.

Usual sector decomposition method divides the integration domain and
find appropriate new variables by an iterative way.
It was found that a simple iterative method may fall into an infinite
loop and \textsl{strategies} are proposed to avoid this problem
by Refs.~\cite{phys:Bogner-Winzier-2008,phys:Bogner-Winzier-2008a}.
Other strategies have been proposed and a practical system has been
constructed by 
Refs.~\cite{phys:Smirnov-Tentyukov-2008,phys:Smirnov-Smirnov-2008,
Smirnov:2009pb}.
Since one has to calculate integration for each sector,
a method is preferable when it produces
less number of sectors.

We propose another method based on the classification of asymptotic
behavior of polynomials around the origin.
The problem is converted to one for convex bodies in Euclidean space
and solved with algorithms developed in computational geometry.
These are deterministic algorithms without iterations.
This talk presents,
based on Ref.~\cite{Kaneko:2009qx},
our basic ideas and show the results of this method.

\section{Sector decomposition}

The procedure of sector decomposition consists of the following
steps:
\begin{enumerate}
\item
Start from Feynman parameter representation of a loop integration.

\item
Primary sector decomposition:

The $\delta$-function is integrated and the integration domain
is decomposed such that the singularities appear in the specific
position in each sector.

\item 
Factorization of the integrand:

The sectors obtained in the previous step are decomposed 
into finer ones. 
New variables are found
such that singular parts are factored out.

\item
Separation of poles in terms of $\EUV$:

The regular part of the integrand is expanded in terms of $\EUV$. 
The coefficients of Laurent series in terms of $\EUV$ are obtained.

\item
Integration of coefficients:

The coefficients are now free from IR divergences.
However, they will still include 
multi-dimensional integration.
They will be integrated out by analytic or numerical methods.

\end{enumerate}

\begin{figure}[!ht]
    \begin{center}
    \includegraphics[width=8cm,clip=true]{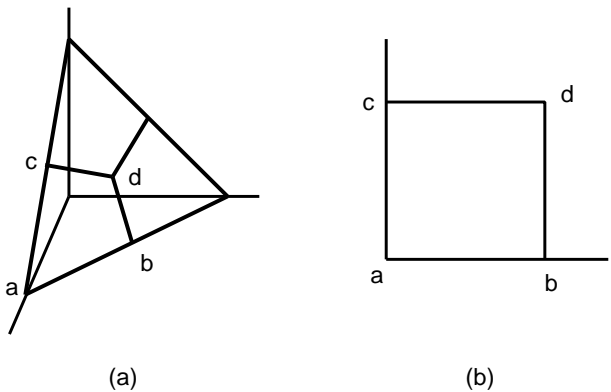}
    \caption{An example of primary sector decomposition in
             3-dimensional case.
             Integration plane is divided into 3 sectors and
             sub-domain $abcd$ in (a) is mapped to 
             a square shown by (b).
             \label{fig:psd} }
    \end{center}
\end{figure}

Feynman parameter representation of a loop integration includes
$\delta$-function which defines the $(N-1)$-dimensional
hyperplane.
The primary sector decomposition integrates over one of integration
variables and decompose the integration domain into $N$ sectors.
An example is shown by Fig.~\ref{fig:psd} for the 3-dimensional case.
By the decomposition of the integration domain and selecting
new variables,
sectors become $(N-1)$-dimensional cubes.
The original boundary of the integration domain is mapped
to a part of the new boundary specified by conditions that some of new
variables are 0.
The resulting expression of the integration for $l$-th sector
becomes:
\begin{equation} \label{sd:gl}
G_l = \int_0^1 d^{N-1} t \; t^{\nu-\VOne}
        \;
        \Polyn{U}_l^{\gamma}(t) \, \Polyn{F}_l^{\beta}(t) ,
\quad
t^{\nu-\VOne} = \prod_{j=0}^{N-1} t_j^{\nu_j-1} ,
\end{equation}
where 
$\Polyn{F}_l(t)$ and $\Polyn{U}_l(t)$ are polynomials of 
variables $t$.
Parameter $\nu_j$ is integer determined by the power of propagators, and
$\VOne$ is an $(N-1)$-dimensional vector whose elements are all 1.
Powers $\beta$ and $\gamma$ are functions of $\EUV$.
If $\Polyn{F}_l(t)$ or $\Polyn{U}_l(t)$ become 0 at the boundary,
and corresponding power $\beta$ or $\gamma$
becomes negative integer for $\EUV=0$, 
the integration diverges.

\section{Geometric method}

We want to express polynomial $\Polyn{F}_l(t)$ in the form:
\begin{equation}
\Polyn{F}_l(t(z)) = C_a z^{b_{a}} \left( 1 + H_{a}(z)\right) 
,\quad 
H_{a}(0) = 0 
,\quad
z^{b_{a}} = \prod_j z_j^{(b_a)_j}
,
\end{equation}
where $z$ is a set of new variable, $H_{a}(z)$ is a polynomial of $z$,
$b_a$ is an integer vector 
and $C_a$ is a constant.
In order to seek such an expression, we consider the following example
taken from one-loop box integration:
\begin{equation} \label{olbex}
\begin{split}
\Polyn{F}_l(t) = - s_{23} t_2 t_3 - s_{12} t_1 - s_4 t_1 t_3
  = t_2 t_3 \left(- s_{23} - s_{12}\frac{t_1}{t_2 t_3}
                  - s_4 \frac{t_1}{t_2}\right).
\end{split}
\end{equation}
If terms $t_1/t_2 t_3$ and $t_1/t_2 \rightarrow 0$ when
$t_2 t_3 \rightarrow 0$, the asymptotic behavior of 
$\Polyn{F}_l$ is determined by the term $- s_{23} t_2 t_3$.
The singular behavior of the integrand around the origin 
is determined by this term.
This condition is satisfied by taking new variables $z$
defined by $t_2 = z_2,\; t_3 = z_3,\; t_1=z_1 z_2 z_3$.
With the range of $z_j$ being limited to $(0, 1)$,
a sub-domain of the integration domain in $t$-space is obtained.
This example shows that 
a term of a polynomial in some sub-domain
will determine asymptotic behavior of the polynomial around the origin.
We call this term \textsl{dominant} in this sub-domain.

In order to see dominant terms more closely around the origin, 
let us change variable
to $y_j = - \log t_j$ or $t_j = e^{-y_j}$.
Monomial $t^c = t_1^{c_1} t_2^{c_2} \cdots t_{N-1}^{c_{N-1}}$
becomes $e^{-(c, y)}$, where $(c,y) = \sum {c_j y_j}$ is
the inner product defined in $(N-1)$-dimensional Euclidean space.
This shows that a monomial corresponds one-to-one to a integer
vector $c$ in this Euclidean space
(let us call this \textsl{power vector}).
Let $Z^{\Polyn{F}_l}$ is the set of power vectors corresponding to the
terms included in a polynomial $\Polyn{F}_l$.
Polynomial $\Polyn{F}_l$ is expressed by:
\begin{equation}
\Polyn{F}_l(t) = \sum_{c \in Z^{\Polyn{F}_l}} a_c t^c 
   = \sum_{c \in Z^{\Polyn{F}_l}} a_c e^{-(c, y)} .
\end{equation}
In order to find the dominant term, let us consider a limit
$\lambda \rightarrow +\infty$ for 
$y_j=\lambda u_j \rightarrow +\infty$ with a fixed non-negative real
vector $u$.
A term $t^b = e^{-(b,y)}$ is dominant in this limit when
$(b, y) \leq (c,y)$ for all $c \in Z^{\Polyn{F}_l}$.
Conversely, let us fix vector $b$ and vary $u$ or $y$.
The term with this power vector $b$ is dominant in the sub-domain
defined by:
\begin{equation} \label{sdexsec}
\Delta_b^{\Polyn{F}_l} := \{ y \in \SetReal_{\geq 0}^{N-1} |
             (c-b, y) > 0, \forall c \in Z^{\Polyn{F}_l}\}.
\end{equation}
This sub-domain forms a convex polyhedral cone.
The polynomial $\Polyn{F}_l$ is expressed as the following:
\begin{equation} \label{sdexexp1}
\Polyn{F}_l(t(y)) = \sum_{b\in Z^{\Polyn{F}_l}} 
   \theta(y \in \Delta_b^{\Polyn{F}_l})
   e^{-(b,y)} \Bigl[
   a_b + \sum_{c \in Z^{\Polyn{F}_l}-\Set{b}} a_c e^{-(c-b,y)}
   \Bigr] .
\end{equation}
For a vector $y \in \Delta_b^{\Polyn{F}_l}$, 
factor $e^{-(c-b,y)} \rightarrow 0$ for
$|y| \rightarrow \infty$, since $(c-b,y) > 0$ holds for all $c$.
This implies that the term $t^b$ is dominant in the sub-domain
$\Delta_b^{\Polyn{F}_l}$.
Thus, asymptotic behavior of $\Polyn{F}_l$ is classified in
terms of these sub-domains $\Set{\Delta_b^{\Polyn{F}_l}}$.

\begin{figure}[!ht]
    \begin{center}
    \includegraphics[width=6cm,clip=true]{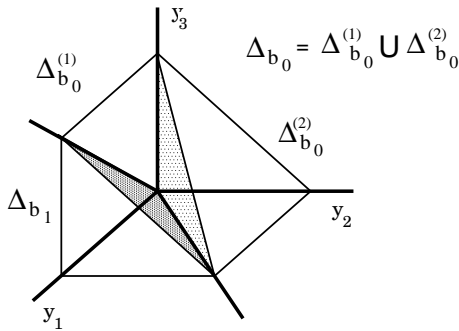}
    \caption{An example of geometric sector decomposition
             \label{fig:sdex} }
    \end{center}
\end{figure}

When this method is applied also to $\Polyn{U}_l$, the singular parts
are factored out from the integration.
Let us return to our example Eq.~(\ref{olbex}) of one-loop box.
In this case
$\Polyn{U}_l$  is $ 1+t_1+t_2+t_3$ and
this expression shows that $\Polyn{U}_l$ does not produce
IR singularity ($\Polyn{U}_l \rightarrow 1$ for $t_i \rightarrow 0$).
With a simple calculation, we obtain from Eq.~(\ref{olbex}):
\begin{equation} \label{sdexexp2}
\begin{split}
t^{b_0} &= t_1, \quad t^{b_1} = t_2 t_3, \quad t^{b_2} = t_1 t_3 ,
\\
Z^{\mathcal{F}_l} &=
  \Set{b_0=(1,0,0),\; b_1=(0,1,1),\; b_2=(1,0,1)} ,
\\
\Delta_{b_0}
 &= \SetIn{x_1(1,1,0)+x_2(1,0,1)+x_3(0,1,0)+x_4(0,0,1)}{
           x_1, x_2, x_3, x_4 \geq 0}
, \\
\Delta_{b_1}
 &= \SetIn{x_1(1,0,0)+x_2(1,1,0)+x_3(1,0,1)}{x_1, x_2, x_3 \geq 0}
, \\
\Delta_{b_2} &= \emptyset .
\end{split}
\end{equation}
The sub-domains are shown in Fig.~\ref{fig:sdex}.
The fact that the last sub-domain is empty 
means that term $t_1 t_3$ never becomes dominant.
The second domain is a triangular cone for which
variables $x$ take simple range of values.
However, the first sub-domain includes four parameters in
3-dimensional space.
When this sub-domain is divided further into two triangular cones,
this redundancy disappears:
\begin{align*}
\Delta_{b_0}
 &= \Delta_{b_0}^{(1)} \cup \Delta_{b_0}^{(2)}
, \\
\Delta_{b_0}^{(1)}
 &= \SetIn{x_1(1,1,0)+x_2(1,0,1)+x_4(0,0,1)}{x_1, x_2, x_4 \geq 0}
, \\
\Delta_{b_0}^{(2)}
 &= \SetIn{x_1(1,1,0)+x_3(0,1,0)+x_4(0,0,1)}{x_1, x_3, x_4 \geq 0}
.
\end{align*}
Now we subsequently change variable from $y_j$ to $x_i$ and then to
$z_i = e^{-x_i}$.
The integration domain for $z$ becomes 3-dimensional unit cube.
Finally we obtain a sector decomposition as the following:
\begin{align*}
\begin{array}{lll}
\hline
\text{sector} & ~\quad\text{variables} & ~\quad Z^{\mathcal{F}_l} \\
\hline
\Delta_{b_0}^{(1)} &
    t_1 = z_1 z_2,\; t_2 = z_1,\; t_3 = z_2 z_4 \quad&
    z_1 z_2 ( - s_{12} - s_{23} z_4 - s_4 z_2 z_4)
\\
\Delta_{b_0}^{(2)} &
    t_1 = z_1,\; t_2 = z_1 z_3,\; t_3 = z_4 \quad&
    z_1 ( - s_{12} - s_{23} z_3 z_4 - s_4 z_4)
\\
\Delta_{b_1} &
    t_1 = z_1 z_2 z_3,\; t_2 = z_2,\; t_3 = z_3 \quad&
    z_2 z_3 ( - s_{12} z_1 - s_{23} - s_4 z_1 z_3)
\\
\hline
\end{array}
\end{align*}
It is easy to calculate Jacobian, which is found to be a monomial of $z$.

For the general case, the decomposition of the integration domain 
in accordance with the classification of the asymptotic behavior,
corresponding to Eqs.~(\ref{sdexsec}) and 
(\ref{sdexexp2}), becomes a problem 
in Euclidean geometry.
It is not so hard to solve this when one
uses \textsl{convex hull algorithms} developed in computational 
geometry.
This decomposition is uniquely determined once a polynomial is given.
It is independent of the choice of an algorithm.
In order to obtain final representation of sector decomposition,
corresponding to the previous table,
it is necessary to triangulate convex polyhedral cones.
This is performed with \textsl{triangulation algorithms}.
The triangulation of convex polyhedral cones is not
determined uniquely and the number of sectors will depend on
the algorithm.

\section{Test implementation and conclusion}

We have made a test implementation of this method.
Our convex hull algorithm is a modified one for convex polyhedral cones
based on the algorithm described in
Ref. \cite{math:Edelsbrunner-1987} for polytopes.
For triangulation of convex polyhedral cones, we have developed
our own algorithm.
The input to our program is given by the package described in
Ref. \cite{phys:Ueda-Fujimoto-2008} and output is passed to the
same package for the separation of poles and integrations of
coefficients.
We have checked our program by comparison with another package
\textsl{qhull} (Ref.~\cite{comp:QHull})
for the convex hull algorithm, 
calculating integration volume for
the triangulation, and comparison of integrated values of several diagrams
with references.
The number of decomposed sectors are shown by Table \ref{ex:tab}.

\begin{table}[!ht]
{\small
\caption{ \label{ex:tab}
Comparison of number of sectors among different methods.
Numbers in column ``H'' are cited from Ref.~\cite{phys:Heinrich-2008}.
Columns 
``A'', ``B'', ``C'', ``S'' and ``X'' indicate corresponding
strategy described in
\cite{phys:Bogner-Winzier-2008} and
\cite{phys:Smirnov-Tentyukov-2008}.
As shown in
\cite{phys:Smirnov-Tentyukov-2008},
``F'' means that the sector decomposition fails and
``M'' means that the memory overflow happened during the sector
decomposition on a 8Gb machine.
The numbers with ``*'' are given by \cite{avsmirnov}.
``This method'' indicates the number of sectors obtained by our method.
``Exponential S.D.'' indicates the number of
sectors before the triangulation.
}
}
\begin{center}
{\small
\begin{tabular}{|l|lllll|l|l|l|}
 \hline
Diagram
    &    A &      B &      C &      S &      X & H & This  & Exponential
\\
    &      &        &        &        &        &   & method & S.D.
\\
 \hline
Bubble
    &   2  &      2 &      2 &      2* &      2 &        &      2 & 2 \\
Triangle
    &   3  &      3 &      3 &      3* &      3 &        &      3 & 3 \\
Box
    &   12 &     12 &     12 &     12  &     12 &        &     12 & 8 \\
\hline
Tbubble
    &   58 &     48 &     48 &     48* &     48 &        &     48 & 36
\\
Double box, \(p_i^2 = 0\)
    &  775 &    586 &    586 &    362  &    293 &    282 &    266 & 106
\\
Double box, \(p_4^2 \neq 0\)
    &  543* &    245* &    245* &    230* &    192* &    197 &    186 & 100
\\
Double box, \(p_i^2 = 0\)
    & 1138 &    698 &    698 &    441* &    395 &        &    360 & 120
\\
~~ nonplanar  & &  & & & & & & \\
D420
    & 8898 &    564 &    564 &    180  &      F &        &    168 & 100
\\
\hline
3 loop vertex (A8)
    & 4617* &   1196* &   1196* &  871* &    750* &    684 &    684 &
240 \\
Triple box
    &     M & 114256  & 114256  &  22657 &  10155 &        &   6568 &
856 \\
 \hline
 \end{tabular}
}
\end{center}
\end{table}

We have proposed a factorization algorithm in sector decomposition.
Our method is based on a classification of asymptotic behavior
of polynomials.
The problem is converted to one for convex bodies in Euclidean space.
In order to find sector decomposition for a given integrand,
we employed algorithms developed in computational geometry.
This method is deterministic and never falls into an infinite loop.
A test implementation shows that the number of decomposed sectors
is less than iterated sector decomposition combined with several
strategies.

\vspace{1em}
\noindent
This work is supported in part by
Ministry of Education, Science, and Culture, Japan
under Grant-in-Aid Nos. 20340063 and 21540286.



\begin{thebibliography}{99}
\bibitem{phys:Binoth-Heinrich-2000}
  T.~Binoth and G.~Heinrich, 
  Nucl.\ Phys.\  B {\bf 585} (2000) 741
  [arXiv:hep-ph/0004013].
  
\bibitem{phys:Binoth-Heinrich-2004}
  T.~Binoth and G.~Heinrich,
  Nucl.\ Phys.\  B {\bf 680} (2004) 375
  [arXiv:hep-ph/0305234].

\bibitem{phys:Binoth-Heinrich-2004a}
  T.~Binoth and G.~Heinrich,
  Nucl.\ Phys.\  B {\bf 693} (2004) 134
  [arXiv:hep-ph/0402265].

\bibitem{phys:Heinrich-2008}
  G.~Heinrich,
  Int.\ J.\ Mod.\ Phys.\  A {\bf 23} (2008) 1457
  [arXiv:0803.4177 [hep-ph]].

\bibitem{phys:Bogner-Winzier-2008}
  C.~Bogner and S.~Weinzierl,
  Comput.\ Phys.\ Commun.\  {\bf 178} (2008) 596
  [arXiv:0709.4092 [hep-ph]].

\bibitem{phys:Bogner-Winzier-2008a}
  C.~Bogner and S.~Weinzierl,
  Nucl.\ Phys.\ Proc.\ Suppl.\  {\bf 183} (2008) 256
  [arXiv:0806.4307 [hep-ph]].

\bibitem{phys:Smirnov-Tentyukov-2008}
  A.~V.~Smirnov and M.~N.~Tentyukov,
  Comput.\ Phys.\ Commun.\  {\bf 180} (2009) 735
  [arXiv:0807.4129 [hep-ph]].

\bibitem{phys:Smirnov-Smirnov-2008}
  A.~V.~Smirnov and V.~A.~Smirnov,
  JHEP {\bf 0905} (2009) 004
  [arXiv:0812.4700 [hep-ph]].

\bibitem{Smirnov:2009pb}
  A.~V.~Smirnov, V.~A.~Smirnov and M.~Tentyukov,
  [arXiv:0912.0158 [hep-ph]].

\bibitem{Kaneko:2009qx}
  T.~Kaneko and T.~Ueda,
  [arXiv:0908.2897 [hep-ph]],
  to appear in Comput. Phys. Commun.

\bibitem{math:Edelsbrunner-1987}
  H.~Edelsbrunner.
  {\em ``Algorithms in Combinatorial Geometry,''}
  Springer, 1987.

\bibitem{phys:Ueda-Fujimoto-2008}
  T.~Ueda and J.~Fujimoto,
  PoS {\bf (ACAT08)} (2008) 120
  [arXiv:0902.2656 [hep-ph]].

\bibitem{comp:QHull}
  Program package \texttt{qhull},
  \verb+http://www.qhull.org/+.

\bibitem{avsmirnov}
  Private communication with A.~V.~Smirnov.

\end{thebibliography}
\end{document}